\documentclass[12pt,aps,preprint,nofootinbib,noshowkeys,noshowpacs,longbibliography,superscriptaddress,tightenlines]{revtex4-2}
\usepackage{amsmath,amsfonts,amsthm,amssymb}
\usepackage{graphics,graphicx}
\usepackage[inline]{enumitem}
\usepackage{mathrsfs}
\usepackage{bm}
\usepackage{color}
\definecolor{darkblue}{RGB}{0,0,196}
\definecolor{darkgreen}{RGB}{0,120,0}

\usepackage{stackengine}

\newcommand{\ubeta}{{\underline{\smash{\beta}}}}

\def\bb{{(b)}}
\def\x{{\bm x}}
\def\p{{\bm p}}
\def\edense{{\mathcal{E}}}
\def\pdense{M}
\def\dpdense{\delta\kern-0.08em M}
\def\dbeta{\delta\kern-0.08em \beta}

\def\noisekernel{{\kappa}}
\def\Eq#1{eq.~(\ref{#1})}
\def\Eqs#1{eqs.~(\ref{#1})}
\def\eq#1{(\ref{#1})}
\def\app#1{App.~\ref{#1}}
\def\Fig#1{Fig.~\ref{#1}}
\def\Figs#1{Figs.~\ref{#1}}
\def\Sect#1{Sect.~\ref{#1}}

\def\beq{\begin{equation}}
\def\eeq{\end{equation}}
\def\st{\begin{equation}}
\def\stp{\end{equation}}
\def\ba{\begin{eqnarray}}
\def\ea{\end{eqnarray}}

\def\tideal{{\mathcal T}}
\def\tvisc{\Pi}

\def\bb1{{(1)}}

\usepackage[colorlinks=true,linktocpage=true,linkcolor=darkblue,citecolor=red,urlcolor=darkblue]{hyperref}

\begin{document}
\preprint{}
 
\title{Relativistic Viscous Hydrodynamics in the Density Frame: Numerical Tests and Comparisons}
    \author{Jay Bhambure}
    \email{jay.bhambure@stonybrook.edu}	
    \affiliation{Center for Nuclear Theory, Department of Physics and Astronomy, Stony Brook University, Stony Brook, New York, 11794-3800, USA}
\author{Aleksas Mazeliauskas}
\email{a.mazeliauskas@thphys.uni-heidelberg.de}
\affiliation{Institute for Theoretical Physics, University of Heidelberg, 69120 Heidelberg, Germany}
    \author{Jean-Fran\c{c}ois Paquet}
    \email{jean-francois.paquet@vanderbilt.edu}
    \affiliation{Department of Physics and Astronomy, Vanderbilt University, Nashville TN 37240}

    \author{Rajeev Singh}
    \email{rajeev.singh@e-uvt.ro}
    \affiliation{Department of Physics, West University of Timisoara, Bd.~Vasile P\^arvan 4, Timisoara 300223, Romania}
    \author{Mayank Singh}
    \email{mayank.singh@vanderbilt.edu}
    \affiliation{Department of Physics and Astronomy, Vanderbilt University, Nashville TN 37240}
    \author{Derek Teaney}
    \email{derek.teaney@stonybrook.edu}
    \affiliation{Center for Nuclear Theory, Department of Physics and Astronomy, Stony Brook University, Stony Brook, New York, 11794-3800, USA}
\author{Fabian Zhou}
\email{f.zhou@thphys.uni-heidelberg.de}
\affiliation{Institute for Theoretical Physics, University of Heidelberg, 69120 Heidelberg, Germany}
	\date{\today} 
	\bigskip
\begin{abstract}

We conduct a numerical study of relativistic viscous fluid dynamics in the Density Frame for one-dimensional fluid flows. The Density Frame is a formulation of relativistic viscous hydrodynamics that is first-order in time,  requires no auxiliary fields,  and has no non-hydrodynamic modes. We compare our results to QCD kinetic theory simulations and find excellent agreement within the regime of applicability of hydrodynamics. Additionally, the Density Frame results remain well-behaved and robust near the boundary of applicability. We also compare our findings to the second-order-in-time hydrodynamic theory developed by Bemfica, Disconzi,  Noronha, and Kovtun (BDNK) and a well-known M\"uller-Israel-Stewart-type hydrodynamics code,  MUSIC,  which is commonly used to simulate heavy-ion collisions.

\end{abstract}
     
\date{\today}
	
\maketitle
\newpage
\tableofcontents

\section{Introduction}

Nuclear collisions at the Relativistic Heavy Ion Collider (RHIC) and the Large
Hadron Collider (LHC) exhibit collective flows, which are accurately described
by relativistic viscous hydrodynamics~\cite{Heinz:2013th}. Comparison between
simulation and experiment have led to increasingly precise estimates of the
shear viscosity of Quantum Chromodynamics (QCD) and have placed strong
constraints on the QCD equation of
state near the crossover transition~\cite{Bernhard:2019bmu,PhysRevLett.126.202301,JETSCAPE,Heffernan:2023kpm}. 

However, the increase in precision together with observed
collective behavior in proton-nucleus collisions and other small systems~\cite{CMS:2015yux,PHENIX:2015idk} have
provided new challenges for viscous hydrodynamics. 
%
%
%
For example, to precisely define the shear viscosity, the hydrodynamic response of the system should be cleanly separated from the non-universal dynamics. 
Similarly, since the number of particles produced in proton-nucleus collisions is limited, fluctuations should be
taken into account by simulating relativistic stochastic fluid
dynamics.  
To address these challenges, this paper and a companion study will examine relativistic viscous fluid dynamics in the ``Density Frame''~\cite{Armas:2020mpr}, which will be defined below.

In the relativistic regime, dissipative fluid dynamics, as
originally developed by Eckart and by Landau and Lifshitz (LL), is
known to suffer from intrinsic
instabilities~\cite{Hiscock:1985zz,GAVASSINO2023137854}. These
instabilities emerge because the divergence of the viscous stress
tensor includes second-order derivatives. When these derivatives
are incorporated into a covariant formalism, the resulting
equations of motion become second order in time, leading to
runaway solutions and other problematic
behaviors~\cite{Hiscock:1985zz,Gavassino:2021owo}. 

To address these issues, several approaches have been proposed.
One prominent solution, developed by M\"uller and by Israel and
Stewart (MIS)~\cite{Muller1967, Israel:1976tn,Israel:1979wp},
introduces additional dynamical variables to the system of
evolution equations. These additional variables are designed to
relax over a collisional timescale, allowing the system to
approach the first-order hydrodynamic behavior described by Landau
and Lifshitz at late times~\cite{Lindblom:1995gp}. 
There are many variants of this
approach~\cite{Israel:1976tn,Israel:1979wp,Geroch:1990bw,OTTINGER1998433,Denicol:2012cn}, and each variant involves some additional fields and parameters.
In linearized analyses, the auxiliary variables give rise to ``non-hydrodynamic''
modes, or modes whose frequency remains finite as the wave number
approaches zero. 
The additional modes complicate the interpretation of the extracted shear viscosity, especially in small systems. 
Nearly all practical simulations of relativistic
dissipative fluids now rely on one of these variants of MIS~\cite{Baier:2007ix,Denicol:2012cn}. We will compare our results to 
the computer code MUSIC~\cite{Schenke:2010nt,Schenke:2010rr,Paquet:2015lta}, which implements a variation motivated by kinetic theory due Denicol, Niemi, Molnar, and Rischke (DNMR)~\cite{Denicol:2012cn}.

A novel reformulation of viscous hydrodynamics was developed by Bemfica, Disconzi, Noronha and subsequently by Kovtun~\cite{Bemfica:2017wps,Kovtun:2019hdm} (BDNK). In their approach no additional
variables are added to Landau and Lifshitz hydrodynamics, but the constitutive
relation is generalized by incorporating all first-order gradients. This expansion includes terms which can be removed using lowest order equations of motion.
BDNK embrace the fact that the relativistic Navier-Stokes equations
are second order in time. In practice, the additional freedom this brings to the
temporal evolution equations functions like the auxiliary
variables and parameters of MIS-type theories. BDN proved that the equations are
strongly hyperbolic and causal~\cite{Bemfica:2017wps}. This motivated
Pandya, Pretorius and Most to study the BDNK system of equations
numerically~\cite{Pandya:2021ief,Pandya:2022pif,Pandya:2022sff}.  We will compare our numerical results to the finite difference BDNK code of \cite{Pandya:2021ief}.

Recently a simpler and more fundamental alternative to
these approaches emerged from an analysis of fluids without boost symmetry\footnote{For example,  a fluid flowing at relativistic speeds over a table is not boost invariant, but is still described by fluid dynamics. Indeed, the stress tensor of the fluid at different fluid velocities (relative to a fixed table) are not related by a Lorentz boost unless the fluid and the table do not  interact.}~\cite{Novak:2019wqg,deBoer:2020xlc,Armas:2020mpr}. The relativistic Navier-Stokes equations in the Density Frame follow as a special case of this analysis~\cite{Armas:2020mpr}, relating terms in the derivative expansion.  In viscous hydrodynamics 
the stress tensor  is divided into an ideal and viscous 
\st
T^{\mu\nu} = \tideal^{\mu\nu}(\beta)   + \Pi^{\mu\nu}  \, , 
\stp
where $\Pi^{\mu\nu}$ is first order in derivatives. 
The ideal stress tensor is 
\st
\tideal^{\mu\nu}(\beta) \equiv \left(e(\beta) + p(\beta)\right)  u^{\mu} u^{\nu} + p(\beta) \, \eta^{\mu\nu} \,  , 
\stp
depends on the inverse temperature and flow velocity,
$\beta_{\mu}(x) \equiv \beta u_{\mu}$.  However, the division of
the stress tensor into ideal and viscous pieces is arbitrary,
since beyond perfect equilibrium the temperature is only
approximately defined. If $\beta_{\mu}$ is shifted 
by a small amount $\beta_{\mu} \rightarrow \beta_{\mu} +
\delta\beta_{\mu}$ changing the hydrodynamic
frame, the viscous stress tensor will change.    Using this
freedom to reparametrize the fields,  the time derivatives in the
viscous stress can be eliminated. The equations of motion are then
strictly first order in time and second order in space and are
stable.  The equations are not relativistically covariant,  but
are invariant under Lorentz transformations followed by a change
of hydrodynamic frame at first order in derivatives. Lorentz
symmetry is presumably restored at infinite order in the
derivative expansion~\cite{Basar:2024qxd}.  
This is similar to many effective
theories used in high energy physics  such as Heavy
Quark Effective Theory~\cite{manohar2000heavy} and Soft Collinear Effective Theory~\cite{Becher:2014oda}, though this similarity has not been formalized.  

The Density Frame is the unique hydrodynamic theory where the
energy and momentum  densities $T^{0\mu}$ on a 
single spatial slice  determine the temperature and flow velocity.
The equations, like ideal hydrodynamics, are first-order
evolution equations for $T^{0\mu}$. Moreover, they have no additional non-hydrodynamic modes or parameters beyond the shear and bulk viscosities and the equilibrium equation of state. 
The Density Frame stress tensor is strictly first order in derivatives  and none 
of the two-derivative corrections of second order hydrodynamics are included. These two-derivative corrections were organized and classified by  Baier, Romatschke, Son, Starinets, and Stephanov (BRSSS)~\cite{Baier:2007ix}. When expanded systematically at late times (after transient non-hydrodynamic modes have decayed), MIS and DNMR hydrodynamics incorporate some of the terms of second (and higher) order hydrodynamics, with specific choices for the transport coefficients~\cite{Baier:2007ix}.

We have previously studied  Density Frame hydrodynamics for the
advection-diffusion equation both theoretically and
numerically~\cite{Basar:2024qxd}.  In particular we analyzed highly
relativistic flows and found excellent agreement with
covariant microscopic approaches in the regime of
applicability of hydrodynamics. At the boundary of applicability, the simulations were well behaved and robust.
This feature,  a requirement for
practical computer codes,  is essentially guaranteed by the
equations of motion, which  become increasingly dissipative at
short-distances.  Although it is not the focus of this paper,
hydrodynamics in the Density Frame places relativistic fluids
squarely in the framework of dissipative stochastic processes.
Indeed, we showed how the stochastic relativistic diffusion
equation can be simulated with a variant of the Metropolis
algorithm~\cite{Basar:2024qxd}. 

Our goal in this paper and a companion study is to extend our work 
on the advection-diffusion equation to the relativistic Navier Stokes equations.  In the companion study~\cite{TheoryPaper}, we analyze  the Density Frame equations of motion in full detail, and
propose an algorithm for stochastic hydrodynamics based on
Metropolis updates.   In this paper we will simulate deterministic
hydrodynamics for a set of one dimensional tests and compare the
results to a variety of other approaches, including the BDNK and
DNMR hydrodynamics. The choice
of tests we analyzed were  guided by a numerical study of the 
BDNK equations of motion by Pandya and Pretorius~\cite{Pandya:2021ief}.  We also performed QCD kinetic theory simulations for
comparison. The numerical simulations of QCD kinetics for a 1+1 dimension expanding system are the first of their kind, and significantly extend earlier work~\cite{Kurkela:2014tea,Keegan:2015avk,Kurkela:2015qoa,Kurkela:2018oqw,Kurkela:2018xxd,Du:2020dvp,Du:2020zqg,Zhou:2024ysb,Boguslavski:2023fdm,Boguslavski:2024kbd}. The numerical details of this contribution will be presented elsewhere~\cite{MazeliauskasZhou}.  
For our purposes kinetic simulations provide a conceptually important
standard -- the kinetic equations are relativistically covariant,
hydrodynamic frame invariant\footnote{Note that the Relaxation Time Approximation (RTA) for kinetic theory is not invariant under change of hydrodynamic frames and is therefore not an appropriate microscopic model for this purpose.}, and can smoothly interpolate between free streaming and hydrodynamics. 

An outline of the paper is the following. In \Sect{sec:densityframe} we provide
a short self-contained introduction to deterministic hydrodynamics in the
Density Frame. In \Sect{sec:numericsoverview} we write down the algorithm used
to simulate the equations of motion.  In \Sect{sec:numericalresults} we compare the Density Frame to the BDNK approach following and extending the
tests studied by Pandya and Pretorius. In \Sect{sec:microscopic} we perform simulations of
QCD kinetic theory~\cite{Arnold:2002zm} and compare its results to the
Density Frame and to the MIS approach in \Sect{sec:MUSIC}. Finally in \Sect{sec:outlook} we
provide a brief summary and outlook.

%

\section{The Density Frame and viscous hydrodynamics}
\label{sec:densityframe}

For simplicity we focus on hydrodynamics in one dimension where 
the evolution variables are the energy and momentum densities 
\st
(T^{00}, T^{0x} ) \equiv (\edense , \pdense^x)\,.
\stp
The spatial stress tensor is $T^{xx}$ and the conservation laws are
\begin{subequations}
\begin{align}
  \partial_t \edense + \partial_x \pdense^x =& 0 \,, \\
  \partial_t \pdense^x + \partial_x T^{xx} =& 0 \,.
\end{align}
\end{subequations}
In order to close the system the stress tensor $T^{xx}$ must be specified. 

The specification of $T^{xx}$ is usually implemented with an intermediate set of parameters,  $\beta_{\mu} = \beta u_{\mu}$, which describe the inverse temperature and four velocity.  
In ideal hydrodynamics, the stress tensor  has 
the functional form
\st
\tideal^{\mu\nu}(\beta) \equiv \left(e(\beta) + p(\beta)\right)  u^{\mu} u^{\nu} + p(\beta) \, \eta^{\mu\nu} \, , 
\stp
where $\eta^{\mu\nu} = (-,+,+,+)$ is the metric tensor.
This equation means  that $\beta_{\mu}$ is determined from the energy and momentum densities:
\begin{subequations}
  \label{eq:densityframedef}
\begin{align}
  \edense =& \tideal^{00}(\beta) \, , \\
  \pdense^x =& \tideal^{0x}(\beta)  \,  , 
\end{align}
\end{subequations}
and subsequently $\beta_{\mu}$ is used to specify the spatial stress,  $T^{xx}  = \tideal^{xx}(\beta)$. In the Density Frame
the algebraic relations in \Eq{eq:densityframedef} define $\beta_{\mu}(x)$ to all orders in the derivative expansion. 

$T^{xx}$ receives viscous corrections of order $\partial_x\beta_x$. We 
show below that the corrections take the form
\st
\label{eq:txx1d}
T^{xx} = \tideal^{xx}(\beta) + \Pi^{xx}\, , \qquad   \Pi^{xx} \equiv  - T\noisekernel^{xxxx}(v) \partial_{(x}\beta_{x)}  \, , 
\stp
where $\kappa^{xxxx}(v)$ takes the form
\st
\label{eq:noisekernelintro}
\noisekernel^{xxxx}(v) =   \frac{4}{3}\frac{\eta}{\gamma^4(1 - v^2 c_s^2)^2} \,.
\stp
Here $\gamma =1/\sqrt{1-v^2}$  and $c_s^2 = dp/de$ and we have neglected the bulk viscosity.   To summarize,  in the next section we will solve the conservation laws with $T^{xx}$ given in \Eqs{eq:txx1d}  and \eqref{eq:noisekernelintro},  and $\beta_{\mu}$ defined through \Eq{eq:densityframedef}.

The functional form in \Eq{eq:noisekernelintro} 
arises from the following steps. In the Landau Frame (which is notated with an underline  $\ubeta_{\mu}$ and $\underline{\Pi}^{\mu\nu}$) the stress tensor takes the form
\begin{subequations}
  \label{eq:Landauedense}
\begin{align}
  \edense =& \tideal^{00}(\ubeta) + \underline{\Pi}^{00}(\ubeta) \, ,   \\
  \pdense^x =& \tideal^{0x}(\ubeta) + \underline{\Pi}^{0x}(\ubeta) \, , 
\end{align}
\end{subequations}
where\footnote{Since the viscous stress is already first order in derivatives, the distinction between $\beta$ and $\ubeta$ is irrelevant in the viscous stress. When changing frames $\ubeta = \beta + \delta\beta$ it is not necessary to include $\delta\beta$ in the viscous stress. } 
\st
\label{eq:Landaustress}
\underline{\Pi}^{\mu\nu} =  -2 T \eta \left[\Delta^{(\mu\rho} \Delta^{\nu) \sigma} - \frac{1}{3} \Delta^{\mu\nu} \Delta^{\rho \sigma}  \right] \partial_{(\rho} \beta_{\sigma)}.
\stp
Here $\Delta^{\mu\nu} = \eta^{\mu\nu} + u^\mu u^\nu$ is the spatial projector. The intermediate parameters $\ubeta(x)$  are redefined,  $\ubeta^{\mu} \equiv  \beta^{\mu} + \delta \beta^{\mu}$, and the ideal stress $\tideal^{0\mu}(\beta + \delta\beta)$ in \eqref{eq:Landauedense}  is expanded to first order in $\delta \beta_\mu$.  $\delta\beta_{\mu}$ is adjusted to cancel  $\underline{\Pi}^{00}$ and $\underline{\Pi}^{0x}$. 
However,  $\delta \beta_{\mu}$ enters into the spatial stress, $T^{xx} =\tideal^{xx}(\beta+\delta\beta) + \underline{\Pi}^{xx}$. 
Expanding $\tideal^{xx}(\beta + \delta \beta)$ to first order,  and using the ideal equations of motion
to replace the time derivatives in \eqref{eq:Landaustress} with spatial derivatives  produces \eq{eq:txx1d} with $\kappa^{xxxx}(v)$ defined in \eqref{eq:noisekernelintro}.  These steps are detailed in \app{app:dfdetails}. 

Next we will analyze entropy production using the Density Frame equations of motion. The entropy current in ideal hydrodynamics  is 
\st
 S^\mu= (S, S v^x) = s(\beta) u^\mu \, , 
\stp
where the intermediate parameters $\beta_{\mu}$ are defined by \eqref{eq:densityframedef}.  This definition of $S\equiv s(\beta) u^0$ (with $\beta_\mu$ defined by $\edense$ and $\pdense$) remains valid in the Density Frame, but the spatial current is modified.  Using the equations of motion it is easy to show  that
\st
\partial_t S + \partial_x (S v^x - \beta_x \Pi^{xx})  =  \kappa^{xxxx}(v) \left(\partial_x\beta_x\right)^2 \, .
\stp
Thus,  it is important that $\kappa^{xxxx}(v)$ in \eqref{eq:noisekernelintro} is a non-negative constant. 

In our companion paper we present the equations of motion in  3+1D including the shear and bulk viscosities~\cite{TheoryPaper}.

\section{Numerical tests in 1+1 dimensions}
\label{sec:numerics}

\subsection{Overview}
\label{sec:numericsoverview}
We will use operator splitting together with an Implicit-Explicit (IMEX) time integrators~\cite{ASCHER1997151,Pareschi2005,Constantinescu} to
evolve the system. The variables
are denoted generically\footnote{For clarity, we will drop the directional superscript in this section,  $\pdense \equiv \pdense^x$.  } $V \equiv(\edense, \pdense)$ and evolve schematically as
\st
  \dot V = f(V) + g(V)\,,
\stp
where $f(V)$ and $g(V)$ represent the ideal and viscous hydrodynamic evolution respectively.
The ideal evolution $f(V)$ is strongly non-linear and will be treated explicitly, while the viscous evolution $g(V)$ will be 
treated implicitly.

In the explicit step, $\dot V = f(V)$,  the variables are evolved over
a time interval $\Delta t$ using the ideal equations of motion:
\begin{align}
  \partial_t \edense + \partial_x \pdense =& 0 \, ,   \\
  \partial_t \pdense + \partial_x \tideal^{xx}(\beta) =& 0 \, . 
\end{align}
For the explicit step we use a standard Kurganov-Tadmor (KT) scheme of ideal hydrodynamics~\cite{KTScheme,KTAstro,Schenke}.
A few further details are given in App~\ref{app:numericaldetails}.

In the implicit step, $\dot V = g(V)$, the variables are  evolved using the viscous equations of motion:
\begin{subequations}
\begin{align}
  \partial_t \edense =&  0 \, ,  \\
  \partial_t \pdense +  \partial_x \tvisc^{xx} =& 0 \, .
\end{align}
\end{subequations}
Abbreviating $\kappa \equiv \kappa^{xxxx}$ to avoid clutter, 
a fully implicit Euler step  consists  of solving the non-linear equations
\begin{subequations}
  \label{eq:FofV}
\begin{align}
  \edense^{n+1} =& \edense^n  \, , \\
  \pdense^{n+1} =& \pdense^n - \Delta t \, \partial_x  \tvisc^{xx} (V^{n+1}) \, .
  \end{align}
\end{subequations}
  Here the viscous strain is spatially discretized
\st
  \label{eq:FofVdetail}
  \partial_x \tvisc^{xx}(V) \equiv 
  -\frac{\kappa(V_{i+1}) + \kappa(V_i) }{2 (\Delta x)^2 }  (\beta_{x, i+1} - \beta_{x,i}) + 
  \frac{\kappa(V_{i}) + \kappa(V_{i-1}) }{2 (\Delta x)^2}  (\beta_{x, i} - \beta_{x, i-1})   \, . 
\stp
Schematically an implicit Euler update takes the form
\st
\label{eq:FofVschematic}
   V^{n+1 } = V^n + \Delta t \, g(V^{n+1}) \, .
\stp
In general \Eq{eq:FofV} and its schematic counter part \eqref{eq:FofVschematic} must be solved by a Newton iterator. 
Combining the explicit and implicit steps sequentially leads to a non-linear Euler IMEX scheme.
Higher order non-linear IMEX schemes can be developed and are studied in \app{app:numericaldetails}.


A simpler alternative to a fully non-linear implicit step is to linearize the problem over a time interval $\Delta t$~\cite{Constantinescu}. In this approach we solve the linearized equation 
\st
\partial_t \, \delta V(t)  =   f(V^n)  +  g(V^n) +  J(V^{n}) \, \delta V\,.
\stp
Here $\delta V(t) = V(t) - V^n$ and $J(V^n) = \partial g/\partial V$ is a time-independent Jacobian matrix.  This leads to the update 
\st
V^{n+1}  = V^n +   \frac{\Delta t \, (f(V^n) + g(V^n)) }{1 - \Delta t J(V^n) }\,,
\stp
which means that the linear equations 
\st
(1 - \Delta t J(V^n)) \delta V  = \Delta t \, (f(V^n) + g(V^n)) \,,
\stp
should be solved for the update $\delta V$. 
In the current context
the linearized equations that evolved 
implicitly over time interval $\Delta t$ 
are
\begin{align}
  \partial_t\, 
    \delta \edense =&  - \partial_x \pdense \\ 
  \partial_t\, \dpdense  + \partial_x \left( \kappa(V^n) \partial_x \dbeta_x
 \right)=&  -\partial_x \tideal^{xx}(V^n) - \partial_x
\tvisc^{xx}(V^n)\, .
\end{align}
Here the derivative of $\dbeta_x$ is shorthand for the differences
\st
\Delta x\,  \partial_x \dbeta_x = \chi^{-1}_{\pdense\edense}  \left(\delta \edense_{i+1} -
  \delta \edense_i\right) +  \chi^{-1}_{\pdense\pdense} \left(\dpdense_{i+1} - \dpdense_i\right) \, .
\stp
Here $\chi^{-1}_{\mu\nu}$ is the static thermodynamic susceptibility matrix
(see App.~\ref{app:dfdetails})
\begin{align}
  \chi^{-1}_{\pdense\pdense} =& \left(\frac{\partial \beta_x}{\partial \pdense} \right)_\edense =  \frac{\gamma \beta}{e +p} \left( \frac{1 + 3c_s^2 v^2 }{1 - v^2 c_s^2}\right) \,  ,  \\
  \chi^{-1}_{\pdense\edense} =& \left(\frac{\partial \beta_x}{\partial \edense} \right)_\pdense =  -\frac{\gamma \beta}{e +p} v_x \left( \frac{1 + 2 c_s^2  + c_s^2 v^2 }{1 - v^2 c_s^2}\right) \, ,
\end{align}
and is evaluated at the appropriate midpoint as in \eq{eq:FofVdetail}.  
We have found that the discretized linear equations are easily solved with just a few sweeps of Bi-Conjugate Stabilized
algorithm (BiCStab) using a Jacobi
pre-conditioner~\cite{trefethen1997numerical,press2007numerical,bueler2020petsc}.
Other linear solvers such as GMRES~\cite{trefethen1997numerical} produced
similar results. 

After implementing the linearized implicit solver in {\tt PETSc}, we adopted the Extrapolated IMEX approach of \cite{Constantinescu} to achieve third order accuracy in the time evolution.

\subsection{Numerical results}
\label{sec:numericalresults}

In the current section we will develop a sequence of numerical tests to
investigate viscous hydrodynamics in the Density Frame. We are guided by a
similar numerical study completed by Pandya and Pretorius~\cite{Pandya:2021ief} of the viscous hydrodynamics using the BDNK setup.  
Indeed,  all of the BDNK numerical results shown here are based on the computer code from this work~\cite{PandyaGithub}. 

Our numerical results are based on a conformal equation of state where $p = e/3$. 
The relation between temperature and energy density is given by\footnote{ Our choice for $e_0$ is marginally different from \cite{Pandya:2021ief}, where $e_0$ was taken to be $10$.  Our choice is motivated by a desire to compare to kinetic theory simulations of a gluon plasma in \Sect{sec:microscopic}. } 
\st
e= e_0 T^4\,,   \qquad  e_0 =  \frac{\nu \pi^2 }{30} \simeq 5.26379 \,,
\stp 
where we chose $\nu = 16$,  corresponding to an ideal gas of gluons with $N_c=3$.  
As the system is conformal, the choice of units is arbitrary. For definiteness, 
we will set ${\rm GeV}$ as our base unit of energy,  defining ${\rm GeV} = \hbar = c = 1$  as a set of computer units. All graphs are presented in these units, though in the text we will indicate ${\rm GeV}$,  and occasionally $c$,  for clarity.

\subsubsection{Smooth data}\label{sec:smooth}
As a first test,  {\tt test1}, we initialized the system with a Gaussian initial condition
\st
\label{eq:test1ic}
e = A e^{-x^2/L^2} + \delta .
\stp
Here we have set $A=0.48\, {\rm GeV}^{4}$, $L=5\,{\rm GeV}^{-1}$ and $\delta=0.12\,{\rm GeV}^4$,  closely related to the test considered in \cite{Pandya:2021ief}. A measure of the mean free path to system size
can be taken from the maximum of the initial Gaussian profile
\begin{align}
  \frac{\ell_{\rm mfp}}{L} 
  \equiv&  \left. \left(\frac{\eta}{s T c_s L } \right) \right|_{t=0,x=0}   \\
\simeq&  1.0 \left( \frac{4\pi \eta/s}{20} \right)
\left(\frac{A}{0.48 \, {\rm GeV}^4} \right)^{1/4}  \left(\frac{5 \, {\rm GeV}^{-1} }{L}  \right).
\label{eq:Knudsen}
\end{align}
In passing to the second line we used the speed of sound $c_s^2=1/3$ and the relation 
between the temperature and energy density,   $e = e_0 T^4$.  We 
have evolved the system until $t =  50.0\,{\rm GeV}^{-1} \simeq 5.8  L/c_s$.

\begin{figure}
  \centering
  \begin{minipage}[c]{0.49\textwidth}
  \includegraphics[width=1.0\textwidth]{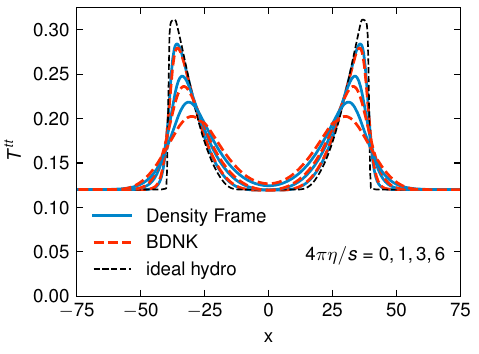}
  \end{minipage}
  \begin{minipage}[c]{0.49\textwidth}
  \includegraphics[width=1.0\textwidth]{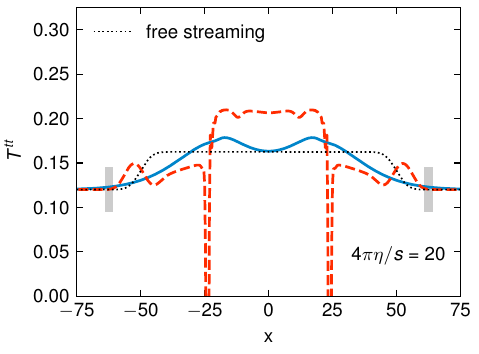}
  \end{minipage}
  \caption{(a) and (b) A comparison between the Density Frame and BDNK for {\tt test1}, a Gaussian initial profile given in \eqref{eq:test1ic}
    with moderate amplitude and thus moderate relativistic effects at,  $t=50.0\,{\rm GeV}^{-1}\simeq 5.8 L/c_s$.  
    (b) The same as in (a) but for larger shear viscosity $4\pi \eta/s=20$.  The grey bands indicate 
    an approximate causality edge, $|x| = 2.5 \, L +  c\, t$, i.e. the perturbations should approximately vanish outside of the grey bands.
  }
  \label{fig:smoothtest1}
\end{figure}

As seen in \Fig{fig:smoothtest1}(a) there is an excellent agreement between the
BDNK and Density Frame  for small shear viscosity, which provide a controlled correction to ideal hydrodynamics (the thin dashed line). 
As the shear viscosity
becomes larger, the differences between the two approaches becomes evident, and these differences are consistent with second order derivative corrections (see App.~\ref{app:MUSICcomparisons} for further discussion).

For $\eta/s\simeq 20/4\pi$ the mean free path to system size (given in \Eq{eq:Knudsen}) becomes of
order unity and in this regime BDNK  develops an oscillatory structure seen in
\Fig{fig:smoothtest1}(b), which is not seen in the Density Frame simulation. Oscillations of this sort were reported previously -- see Fig.~2(c) of ~\cite{Pandya:2021ief}.  

To analyze this regime,  we first note that the initial conditions are Gaussian  with a width parameter  of $L=\sqrt{2} \sigma$, and thus the initial perturbation is approximately contained in the compact interval $|x| < 2.5\,L$. In a Lorentz covariant theory,  the solution should approximately vanish for $|x| > 2.5\, L +  c\, t$ at subsequent times.  We have indicated the approximate causality edge by the grey bands in \Fig{fig:smoothtest1}(b). We also have indicated the stress tensor from a free streaming Bose gas by the dotted line, which is the limit $\ell/L \rightarrow \infty$ in kinetic theory (see \Sect{sec:microscopic}).

\begin{figure}
  \centering
  \begin{minipage}[c]{0.49\textwidth}
  \includegraphics[width=1.0\textwidth]{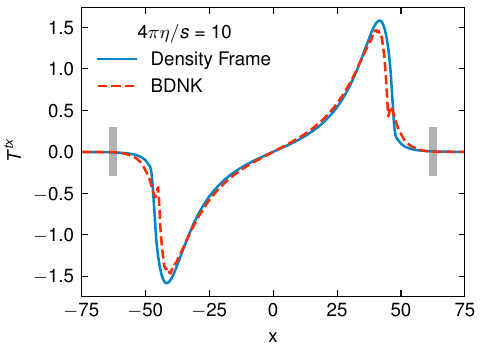}
  \end{minipage}
  \begin{minipage}[c]{0.50\textwidth}
  \includegraphics[width=1.0\textwidth]{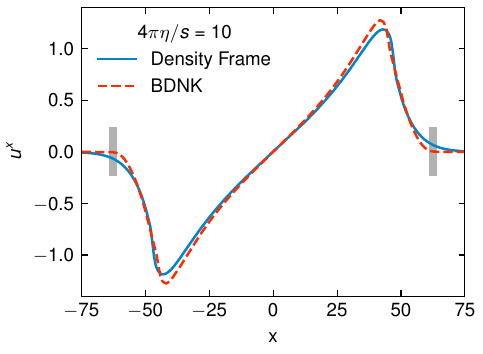}
  \end{minipage}
  \caption{(a) 
    A comparison of the momentum density $T^{tx}$ for the Density Frame and BDNK in  {\tt test2}  at $t=50.0\,{\rm GeV}^{-1}$ and $4\pi\eta/s=10$. The BDNK model with the adopted finite difference scheme begins to develop spurious oscillations as in \Fig{fig:smoothtest1} for $|x|\simeq 50$. For larger values of the shear viscosity the BDNK code is unable to complete the test.
    (b) A similar comparison for the flow velocity $u^x$. The grey lines indicate the approximate causality edge, i.e. Lorentz invariant solutions
    should vanish outside of the grey lines.
  }
  \label{fig:smoothtest2}
\end{figure}

The observed oscillations in the BDNK theory are related to its Lorentz invariant character, which requires the equations of motion to be second order in time.
Indeed, for initial conditions with compact support,  perturbations in BDNK must strictly vanish outside of a compact domain at later times. For the approximately compact initial conditions considered here, this property effectively enforces a theta function in coordinate space,   with  $T^{\mu\nu}$ vanishing beyond the grey bands in \Fig{fig:smoothtest1}(b).  For the second-order BDNK equations,  this constraint  leads to  Gibbs's oscillations in the computational domain. 
While the very sharp spikes could be ameliorated by a different discretization scheme (perhaps with the finite volume scheme of \cite{Pandya:2022pif}),  the remaining oscillations are a consequence of the second-order equations of motion, which are oscillatory at high frequency.
By contrast, the Density Frame
has an exponentially small  tail outside of the (approximate) causality domain, but is generally diffusive in character. This reflects the fact that dynamics of the Density Frame becomes increasingly dissipative in the UV and the equations are first order in time. 
As discussed below, similar features are seen  in Figs.~\ref{fig:smoothtest2}, \ref{fig:smallstep}, and \ref{fig:largestep}.

The first test,  {\tt test1},  is not very relativistic.
Thus,  we  have considered a second initial condition,  {\tt test2},  in the same class, by increasing the amplitude by a factor of $20$ and lowering the baseline by a factor of two,  $A = 9.6 {\rm GeV}^4$ and $\delta = 0.06\, {\rm GeV}^4$. This marginally changes the mean free path to system size, but increases the $\gamma$ factor. Indeed, in ideal hydrodynamics the gamma factor with this initial condition
becomes of order  $\gamma \simeq 1.6$ at $t\simeq 50. {\,\rm GeV^{-1}}\simeq 5.8\,L/c_s$. 
For small viscosity $\eta/s=1,3,6$ the agreement between the Density Frame and
BDNK is similar to \Fig{fig:smoothtest1}(a) (not shown). Starting at
$4\pi\eta/s\simeq 10$, the BDNK again develops an oscillatory structure in
the stress tensor as seen in $T^{tx}$ shown in \Fig{fig:smoothtest2}(a).  These oscillations become increasingly severe with increasing  shear viscosity and for
$4\pi\eta/s=12$, the BDNK solution obtained with the finite difference scheme from \cite{Pandya:2021ief} is unable to complete the test. 
\Fig{fig:smoothtest2}(b) shows the velocity $u^x$ in the simulation. As anticipated, the fluid velocity vanishes outside
of a compact domain in the BDNK formulation, which has engendered spurious oscillations in the rest of the computational domain.  By contrast, the fluid velocity in the Density Frame has an exponentially decreasing tail outside of the causality domain (\Fig{fig:smoothtest2}(b)), which nevertheless contributes negligibly to the momentum $T^{tx}$ (\Fig{fig:smoothtest2}(a)).
\subsubsection{Shock Tube tests}

The next class of test problems studied discontinuous initial data. Here at 
time $t=0$ the energy density is separated by a wall 
\st
   e_0(x)  =  \begin{cases}
     e_L    & x < 0  \\
     e_R    & x \geq 0 
   \end{cases}.
   \label{eq:shockwall}
\stp
This is evolved in time and a rich shock structure develops. For ideal hydrodynamics a concise summary in the relativistic domain is given in \cite{SCHNEIDER199392,RISCHKE1995346}. 

\begin{figure}
  \centering
  \begin{minipage}[c]{0.49\textwidth}
    \includegraphics[width=1.0\textwidth]{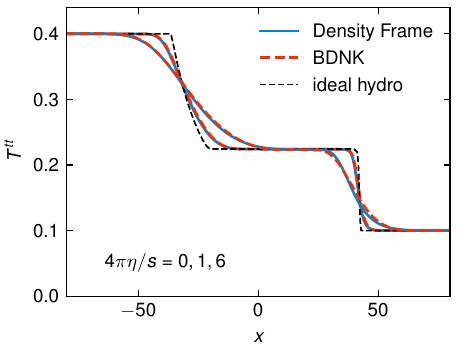}
  \end{minipage}
  \begin{minipage}[c]{0.49\textwidth}
    \includegraphics[width=1.0\textwidth]{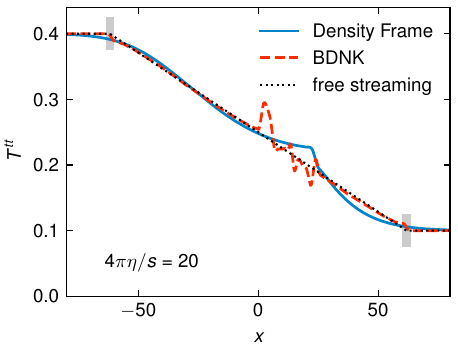}
  \end{minipage}
  \caption{(a) A comparison between the Density Frame, BDNK, and ideal hydrodynamics for a shock tube test ({\tt test3}) evolved to time $t=61.875 \,{\rm GeV}^{-1}$ for $4\pi \eta/s=0,1,6$ (see \Eq{eq:shockwall} for context). 
    The maximum flow velocity is relatively small in this test $u^x\simeq 0.3$.  
    (b) The same as (a) but for $4\pi\eta/s=20$. The grey bands are centered at the Lorentz causality edge. The free streaming solution is discussed in the text, and linearly interpolates between $e_L$ and $e_R$. 
  }
  \label{fig:smallstep}
\end{figure}

We define a third test,  {\tt test3}, by taking the same initial  data as Pandya and
Pretorius with $e_{L}=0.48\,{\rm GeV}^4$ and $e_{R}=0.12\, {\rm GeV}^4$~\cite{Pandya:2021ief}. 
This  is a not a large jump and the maximum flow velocity is only $u^{x}\simeq 0.3$. We will describe a more relativistic test below after studying this initial test. As discussed previously, the shear viscosity sets a typical mean free path 
\begin{align}
  \label{eq:mfpdef}
  \ell_{\rm mfp}
  \equiv&  \left(\frac{\eta}{s T c_s  } \right)
     = 3.6\,{\rm GeV}^{-1} \,  \left( \frac{4\pi \eta/s}{10} \right)
     \left(\frac{0.2\, {\rm GeV}^4}{e} \right)^{1/4}  \, . 
\end{align} 
The shear viscosity produces diffusion and,  over a time $t$,   a  sharp front
is expected to spread by a squared distance of order $(\Delta x)^2 \simeq (\eta/sT) t$, leading to an  estimate for the shock width
\st
\label{eq:diffusivedef}
\sqrt{\Delta x^2 } =  11. {\rm GeV}^{-1}  \left( \frac{4\pi \eta/s}{10 } \right)^{1/2}
\left(\frac{0.2\,  {\rm GeV}^4}{e} \right)^{1/8}  \left( \frac{t}{60 \, {\rm GeV}^{-1}} \right)^{1/2}  \, . 
\stp
\Fig{fig:smallstep}(a) shows how the shear viscosity smooths out the shock
front for small and moderate values of the shear viscosity. The viscous solution is consistent between the Density Frame and BDNK approaches, 
and the viscous diffusion is consistent with the estimate  given in  \eq{eq:diffusivedef}.
For larger viscosities a pattern similar to the smooth test,  {\tt test1},   is
seen in \Fig{fig:smallstep}(b). Specifically, the BDNK approach is strictly causal and this feature
leads to Gibbs oscillations in the rest of the computational domain. 
By contrast the 
Density Frame becomes increasingly viscous (as opposed to oscillatory) in this
regime. Note, however, that the Density Frame solution does have tails which extend beyond the limits imposed by Lorentz causality. Both approaches should be  compared to the free streaming solution discussed in \Sect{sec:microscopic} which has compact support in the causal region,  and linearly interpolates between the $e_L$ and $e_R$.  

\begin{figure}
  \centering
  \begin{minipage}[c]{0.49\textwidth}
    \includegraphics[width=1.0\textwidth]{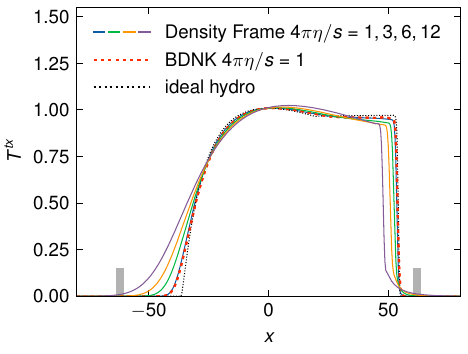}
  \end{minipage}
  \begin{minipage}[c]{0.49\textwidth}
    \includegraphics[width=1.0\textwidth]{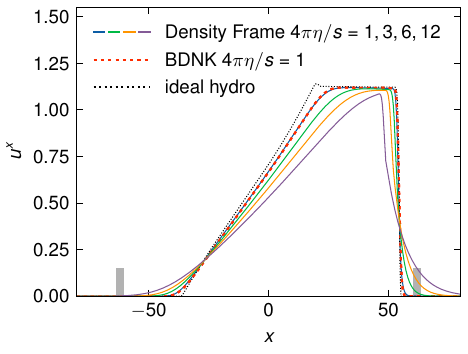}
  \end{minipage}
  \caption{(a) A comparison between the Density Frame, BDNK, and ideal hydrodynamics for a shock tube test ({\tt test4}) with $e_L=4.8\, {\rm GeV}^{4}$ and $e_R = {0.06}\,{\rm GeV}^4$ evolved to time $t=61.875\, {\rm GeV}^{-1}$ for 
    a range of shear viscosities.  For $4\pi\eta/s=3$ and above the BDNK code becomes unstable and is unable to complete the test. See text for further discussion.
    (b) The same as (a) but for the flow velocity $u^x$.  The grey bands indicate the Lorentz causality edge. See text for further discussion.
  }
  \label{fig:largestep}
\end{figure}

The third test, {\tt test3}, is not very relativistic. Thus, 
we consider a fourth test, {\tt test4}, increasing the amplitude
of the shock tube problem, taking $e_L=4.8\, {\rm GeV}^4$ and $e_R = 0.06\,{\rm GeV}^4$. This leads
to a relativistic flow with  $u^x\simeq 1$.  In \Fig{fig:largestep} (a) and (b)
we show the momentum $T^{tx}$ and flow velocity $u^x$ at time $t=61.875\,{\rm GeV}^{-1}$ resulting from the initial conditions of {\tt test4}. 

For the smallest shear viscosity,  $4\pi\eta/s=1$,  the Density Frame
and BDNK agree nicely. However, for $4\pi\eta/s=3$ and higher the BDNK code
becomes unstable and is unable to complete the {\tt test4}. As in {\tt test3}, the instabilities arise precisely when the viscous diffusive front approaches the causality edge, leading to oscillations.  
By contrast, for $4\pi\eta/s=3$ the Density Frame has an small unphysical value beyond the limits set by causality but the solution remains well behaved.  This tail is much less evident in the momentum density $T^{tx}$ (\Fig{fig:largestep}(a)) as opposed to the flow velocity $u^x$ (\Fig{fig:largestep}(b)).

\section{QCD Kinetic Theory}
\label{sec:microscopic}

In this section, we solve 1+1 dimensional expansion problem using relativistic QCD kinetic theory~\cite{Arnold:2002zm}. For simplicity, we will restrict ourselves to Yang-Mills sector with gluon
 phase-space distribution function $f(t,\x,\p)$, which obeys a relativistically covariant Boltzmann equation
\begin{equation}
   p^\mu \partial_\mu f(t,\x,\p) = -C[f]\,. \label{eq:EKTfull}
\end{equation}
The collision kernel $C[f]$ contains leading order elastic 2$\leftrightarrow$2  and inelastic 1$\leftrightarrow$2 scattering processes~\cite{Arnold:2002zm}, 
\begin{equation}
    C[f] = C_{2\leftrightarrow2}[f] + C_{1\leftrightarrow2}[f].
\end{equation}
Here $C_{2\leftrightarrow2}[f]$ and $C_{1\leftrightarrow2}[f]$ are Lorentz invariant multi-dimensional collision integrals with a single external parameter -- the coupling constant $\lambda = g^2N_c$, where $N_c=3$. As $\lambda\to 0$, kinetic theory smoothly approaches  free streaming, while for finite $\lambda$ kinetic theory equilibrates and \Eq{eq:EKTfull} reproduces viscous hydrodynamics.

Solutions of \Eq{eq:EKTfull} in homogeneous and boost-invariant systems have been used in many previous publications~\cite{Kurkela:2014tea,Keegan:2015avk,Kurkela:2015qoa,Kurkela:2018oqw,Kurkela:2018xxd,Du:2020dvp,Du:2020zqg,Zhou:2024ysb,Boguslavski:2023fdm,Boguslavski:2024kbd}.  Thanks to boost-symmetry, the problem can be reduced to 1+0 spatial and 2 momentum dimensions~\cite{Baym:1984np}.
Here, we extend previous work and solve \Eq{eq:EKTfull} for systems that are isotropic and homogeneous in the transverse plane but have no particular symmetry in the third direction, i.e. 1+1 spatial and 2 momentum directions. In this case \Eq{eq:EKTfull} simplifies to
\begin{equation}
    \left(\partial_{t} + v^x \partial_x\right) f(t,x,p,v^x) = -\frac{1}{p}C[f],\label{eq:EKT1d}
\end{equation}
where $v^x$ is the projection of particle velocity along the $x$-axis, i.e., $v^x=p^x/p$ and $p=p^0=|\p|$. 
We discretize linearly in $x$, $p$, and $v^x$ with a  grid of size $(N_x, N_p, N_{v^x}) = (150, 80, 120)$ and intervals of $x\in [-75, 75]$, \, $p \in [0.05,20]$,  and $v^x \in [-1,1]$. 
In each time step of size $\Delta t=1.5\times 10^{-3}$, the collision kernels are evaluated using Monte-Carlo sampling. 
We refer to previous publications for details on the numerical implementation of the collision kernels~\cite{Kurkela:2014tea,Keegan:2015avk,Kurkela:2015qoa,Kurkela:2018oqw,Kurkela:2018xxd,Du:2020dvp,Du:2020zqg,Zhou:2024ysb,Boguslavski:2023fdm,Boguslavski:2024kbd}. A typical simulation takes around one day of runtime with 200 CPU threads.
The full implementation details will be presented in the upcoming publication~\cite{MazeliauskasZhou}.

We initialize the gluon distribution function at $t=0$ with a Bose-Einstein distribution
\begin{equation}\label{therm_distr}
    f_{\mathrm{th}}(p) = \frac{1}{e^{p/T(x)}-1},
\end{equation}
where $T(x)$ is a spatially dependent temperature. We choose $T(x)$ to reproduce the initial energy density profile of hydrodynamic simulations in \Sect{sec:smooth}.
The energy density at the initial moment is given by
\begin{equation}\label{eq:e_landau_matching}
    e(x)=T^{tt}(x)=\nu_g\int\frac{d^3 p}{(2\pi)^3} p f_{\mathrm{th}}(p)= \frac{\nu_g\pi^2}{30}T(x)^4\, ,
\end{equation}
where $\nu_g=16$ is the color and polarization degrees of freedom for gluons.
We choose $T(x)$ to reproduce the Gaussian initial conditions of tests {\tt test1} and {\tt test2}, and evolve until $t=50\, {\rm GeV}^{-1}$ for different values of the coupling constant $\lambda=5,10,20$.

Different coupling constants correspond to different values of specific shear viscosity, and at small coupling this value  can be systematically computed~\cite{Arnold:2003zc}.
For moderate values of $\lambda$, the exact numerical value of $\eta/s$ depends on the implementation details of the leading order collision kernels~\cite{Boguslavski:2024kbd}.
In particular, in this work we use the full isotropic HTL screening prescription for the elastic scattering matrix elements~\cite{Boguslavski:2023fdm, Boguslavski:2024kbd}.
The couplings $\lambda=5,10,20$ then correspond to $\eta/s=1.48,0.513,0.180$~\cite{Boguslavski:2024kbd}, which we use in hydrodynamic comparisons.

As $\lambda\to 0$, kinetic theory smoothly approaches the free streaming with effective $\eta/s=\infty$.
We note that in the free streaming case $C[f]=0$,  and  \Eq{eq:EKT1d} can be solved analytically: $f(t,x,p,v^x)=f(t_0=0,x-v^x(t-t_0),p,v^x)$. Furthermore, for isotropic initial conditions (which are independent of $v^x$) one can show that energy $T^{tt}$ and momentum $T^{tx}$ densities at later times are given by
\begin{align}
T^{tt}(t,x)
&= \frac{1}{2t}\int^{x+t}_{x-t} dx' T^{tt}(t_0=0,x')\, ,\\
T^{tx}(t,x)
&= \frac{1}{2t^2}\int^{x+t}_{x-t} dx' (x-x') T^{tt}(t_0=0,x')\, ,
\end{align}
i.e., free-streaming spreads the initial energy density uniformly over $[x-t, x+t]$ interval.

\begin{figure}
  \centering
  \begin{minipage}[c]{0.49\textwidth}
    \includegraphics[width=1.0\textwidth]{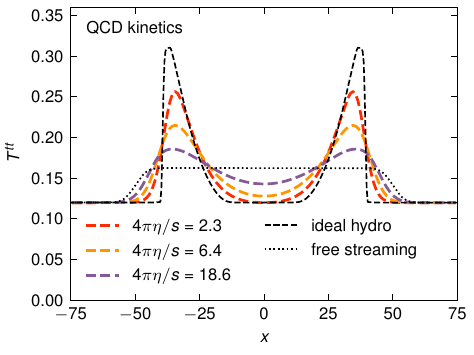}
  \end{minipage}
  \begin{minipage}[c]{0.49\textwidth}
    \includegraphics[width=1.0\textwidth]{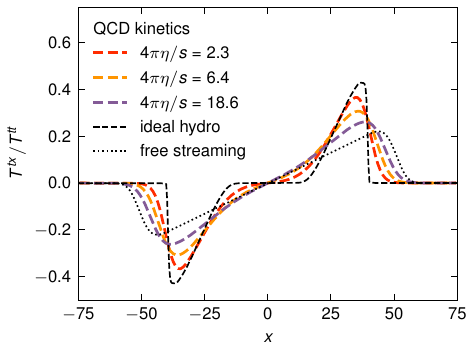}
  \end{minipage}
  \caption{(a) The energy density $T^{tt}$ and (b) the momentum density $T^{tx}$ normalized by $T^{tt}$ of {\tt test1} at $t=50\, {\rm GeV}^{-1}$ in QCD kinetic theory for different couplings, $\lambda=5$, $10$ and $20$, which correspond to $4\pi\eta/s=18.6$, $6.4$ and $2.3$. The dashed and the dotted black line correspond to ideal hydro ($\eta/s = 0$) and free streaming ($\eta/s = \infty$), respectively.
  }
  \label{fig:QCDkinetics1}
\end{figure}

\begin{figure}
  \centering
  \begin{minipage}[c]{0.49\textwidth}
    \includegraphics[width=1.0\textwidth]{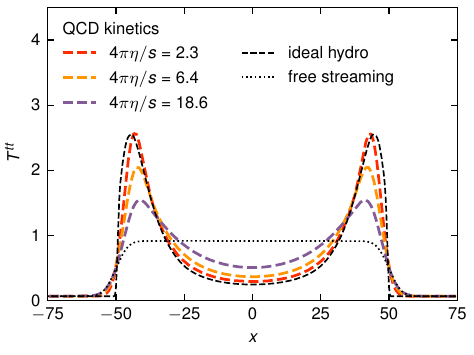}
  \end{minipage}
  \begin{minipage}[c]{0.49\textwidth}
    \includegraphics[width=1.0\textwidth]{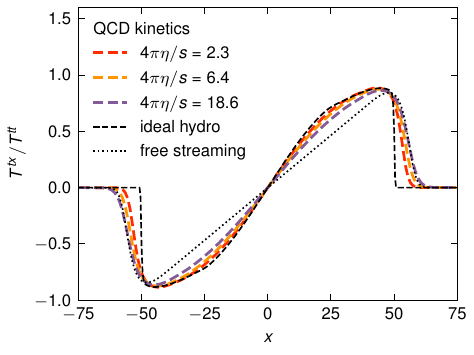}
  \end{minipage}
  \caption{(a) The energy density $T^{tt}$ and (b) the momentum density $T^{tx}$ normalized by $T^{tt}$ of {\tt test2} at $t=50\, {\rm GeV}^{-1}$ in QCD kinetic theory for different couplings as in \Fig{fig:QCDkinetics1}. The dashed and the dotted black line correspond to ideal hydrodynamics ($\eta/s = 0$) and free streaming ($\eta/s = \infty$), respectively.
  }
  \label{fig:QCDkinetics2}
\end{figure}

QCD kinetic theory results for the energy and momentum density evolution of {\tt test1} and {\tt test2} are shown in \Figs{fig:QCDkinetics1} and \ref{fig:QCDkinetics2} respectively. For the mostly non-relativistic case in \Fig{fig:QCDkinetics1}, we see that kinetic theory smoothly interpolates between ideal hydro (dashed lines) and free streaming (dotted lines) limits.
In the relativistic case shown in \Fig{fig:QCDkinetics2},  we observe that although in the inner region $|x|<50$ the kinetic theory is close to ideal hydro result (especially for the largest couplings), in the low-density outskirts $|x|>50$, kinetic theory is much closer to the free streaming regime. This is because in the inner high-density regions, the viscous corrections are small compared to the equilibrium temperature, i.e., $\eta/s\partial_x v/T\ll 1$, but this is not the case for $|x|>50$.
This demonstrates that, in this example, the full dynamics requires the simultaneous description of free streaming and hydrodynamic behavior.
In particular, the particles at the causality edge have never scattered, and thus the causality edge should not be described with hydrodynamics.

\section{Comparison between the Density Frame, M\"uller-Israel-Stewart-type hydrodynamics and QCD Kinetic Theory}
\label{sec:MUSIC}

\begin{figure}
  \centering
  \begin{minipage}[c]{0.49\textwidth}
    \includegraphics[width=1.0\textwidth]{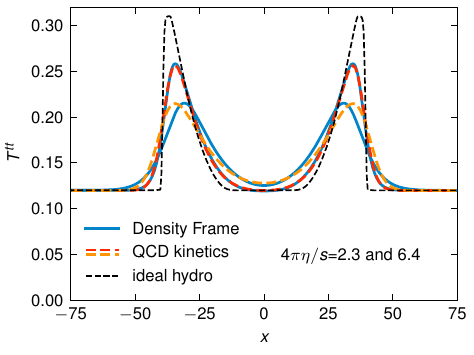}
  \end{minipage}
  \begin{minipage}[c]{0.49\textwidth}
    \includegraphics[width=1.0\textwidth]{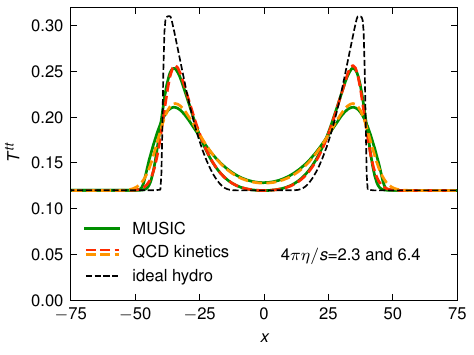}
  \end{minipage}
  \begin{minipage}[c]{0.49\textwidth}
    \includegraphics[width=1.0\textwidth]{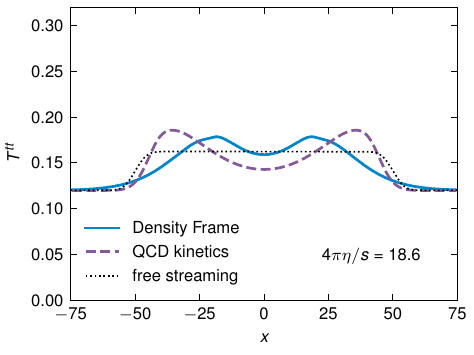}
  \end{minipage}
  \begin{minipage}[c]{0.49\textwidth}
    \includegraphics[width=1.0\textwidth]{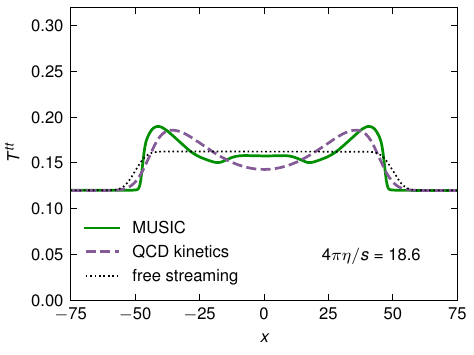}
  \end{minipage}
  \caption{The upper panels show the energy density $T^{tt}$ of {\tt test1} at $t=50\, {\rm GeV}^{-1}$ in QCD kinetic theory compared to (a) the Density Frame and (b) second order hydrodynamics (MUSIC) for $4\pi \eta/s=2.3,6.4$. The dashed black line corresponds to ideal hydrodynamics ($\eta/s = 0$). The lower panels show the same but for a larger shear viscosity $4\pi \eta/s=18.6$. The dotted black line corresponds to free streaming ($\eta/s = \infty$).
  }
  \label{fig:music_test1}
\end{figure} 

\begin{figure}
  \centering
  \begin{minipage}[c]{0.49\textwidth}
    \includegraphics[width=1.0\textwidth]{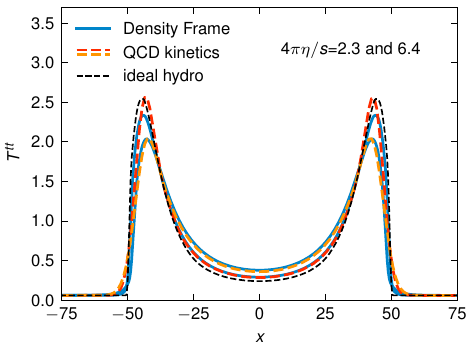}
  \end{minipage}
  \begin{minipage}[c]{0.49\textwidth}
    \includegraphics[width=1.0\textwidth]{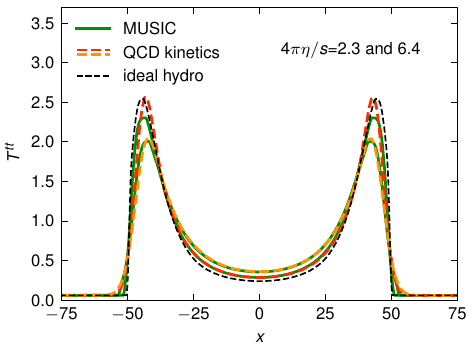}
  \end{minipage}
  \begin{minipage}[c]{0.49\textwidth}
    \includegraphics[width=1.0\textwidth]{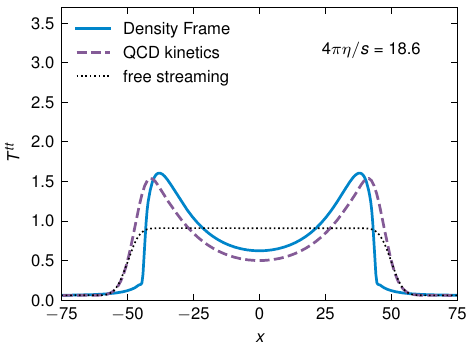}
  \end{minipage}
  \begin{minipage}[c]{0.49\textwidth}
    \includegraphics[width=1.0\textwidth]{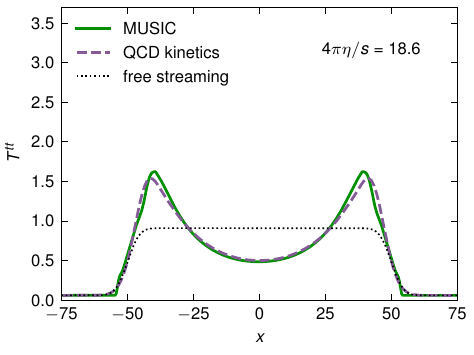}
  \end{minipage}
  \caption{The upper panels show the energy density $T^{tt}$ of {\tt test2} at $t=50\, {\rm GeV}^{-1}$ in QCD kinetic theory compared to (a) the Density Frame and (b) second order hydrodynamics (MUSIC) for $4\pi \eta/s=2.3,6.4$. The dashed black line corresponds to ideal hydrodynamics ($\eta/s = 0$). The lower panels show the same but for a larger shear viscosity $4\pi \eta/s=18.6$. The dotted black line corresponds to free streaming ($\eta/s = \infty$).
  }
  \label{fig:music_test2}
\end{figure} 

Most hydrodynamic simulations of quark-gluon plasma are based on M\"uller-Israel-Stewart-type hydrodynamics~\cite{Muller1967, Israel:1976tn,Israel:1979wp}. Specifically,  most groups use a truncated version of DNMR hydrodynamics~\cite{Denicol:2012cn} or a resummed BRSSS approach~\cite{Baier:2007ix} with varying choices of second-order transport coefficients.
In this section, we compare the DNMR-based MUSIC hydrodynamic simulations~\cite{Schenke:2010nt,Schenke:2010rr,Paquet:2015lta} with both Density Frame hydrodynamics and with the 1+1D QCD kinetic theory results discussed in the previous section. A brief summary of the equations solved in MUSIC can be found in \app{app:MUSICcomparisons}. We highlight that DNMR has non-hydrodynamic modes and second-order transport coefficients, and Density Frame hydrodynamics does not. Consequently, MUSIC's results depend on the choice of second-order transport coefficients, a dependence discussed in \app{app:MUSICcomparisons}. 
We use the typical prescription for the second-order transport coefficients in MUSIC, e.g. $\tau_\pi=5 \eta/(\epsilon+P)$ (see \cite{Denicol:2014vaa,Paquet:2015lta,Ryu:2017qzn} and \app{app:MUSICcomparisons} for details). As in the rest of this work, bulk viscosity is not included.
Since DNMR is derived from kinetic theory, and $\tau_\pi=5 \eta/(\epsilon+P)$ is close to the shear relaxation time in QCD kinetic theory~\cite{York:2008rr,Ghiglieri:2018dgf}, there is a general expectation that MUSIC should outperform Density Frame hydrodynamics in this comparison with kinetic theory.

We first study the same {\tt test1} as in the previous sections.
The specific shear viscosity is again set to $4\pi \eta/s=2.3, 6.4, 18.6$. 
For the lowest shear viscosity ($4\pi \eta/s=2.3$), 
\Fig{fig:music_test1} shows that both Density Frame (top left) and MUSIC (top right) describe the 1+1D QCD kinetic theory result similarly.
On the other hand, if $\eta/s$ is almost tripled ($4\pi \eta/s=6.4$), differences between kinetic theory and Density Frame hydrodynamics can be observed,
while MUSIC  describes the kinetic theory result well. The differences between MUSIC and the Density Frame are consistent with second order terms in the gradient expansion, as can be seen from \Fig{fig:MUSIC_taupi} in App.~\ref{app:MUSICcomparisons} where the second order parameter $\tau_\pi$ is varied. The BDNK simulations describe 
QCD kinetic theory results almost as well the Density Frame (not shown).  
For the largest viscosity, $4\pi \eta/s=18.6$, neither Density Frame hydrodynamics nor MUSIC match the QCD kinetic theory result very accurately, overestimating the result in some regions and underestimating it in others. 

For {\tt test2}, where relativistic expansion effects are larger, the results are shown in \Fig{fig:music_test2}. In this case, Density Frame and MUSIC perform similarly for the two smallest viscosities. For the very largest viscosity, $4\pi \eta/s=18.6$, MUSIC describes well the QCD kinetic theory result, though the result is sensitive to $\tau_\pi$ as can be seen from \Fig{fig:MUSIC_taupi} in App.~\ref{app:MUSICcomparisons}. On the other hand, Density Frame captures the overall shape of the kinetic theory calculations but does not describe it very accurately. MUSIC's performance might again be a sign of its partial ability to capture the non-hydrodynamic modes of kinetic theory.

In summary, for {\tt test1} and {\tt test2}, MUSIC's DNMR implementation somewhat outperforms Density Frame hydrodynamics for one of the three values of shear viscosity tested in comparison to QCD kinetics, although the improvement is not for the same viscosity in the two tests.
Additional numerical studies will be necessary to understand the differences between DNMR and Density Frame hydrodynamics for the fluid geometries and expansion rates simulated in heavy-ion collisions. 
In  \app{app:MUSICcomparisons}  we studied how MUSIC's results depend on the shear relaxation time $\tau_\pi$ (see \Fig{fig:MUSIC_taupi}),  and found that the spread of predictions  is comparable 
to the differences between the Density Frame and 
default MUSIC.
We note that, when it comes to modeling heavy-ion collisions, the relaxation time is not known and the underlying medium is not described by kinetic theory; hence the current tests are not necessarily evidence for either approach's ability to describe strongly-coupled quark-gluon plasma.
Moreover, the additional parameters (second-order transport coefficients) in MIS/DNMR and BDNK make them more flexible models than Density Frame hydrodynamics: it gives them more freedom to describe measurements. Whether this additional flexibility of DNMR and BDNK necessarily translates into additional predictive power for the models is a more difficult question, linked to the nature of the non-hydrodynamic modes of these theories and the actual non-hydrodynamic modes of strongly-coupled quark-gluon plasma.
Nevertheless, Density Frame hydrodynamics, with no non-hydrodynamic mode and consequently fewer parameters, appears to be competitive with hydrodynamic simulations such as MUSIC to describe 1+1D relativistic expansion. We expect it to be an important formalism to understand the role of non-hydrodynamic modes in simulations of heavy-ion collisions.

\section{Outlook}
\label{sec:outlook}

We have investigated the Density Frame approach to relativistic viscous fluids with a series of one-dimensional numerical tests. This framework provides an alternative formulation of relativistic viscous hydrodynamics, distinct from other widely used approaches. While the Density Frame is not explicitly covariant,  it is invariant under the combined action of a boost and a change of hydrodynamic frame to the required order in the derivative expansion.

We compared the Density Frame to QCD kinetic theory, a non-equilibrium covariant microscopic theory.
The Density Frame showed excellent agreement with QCD kinetics within the regime of validity of hydrodynamics (large coupling, or small shear viscosity). 
Moreover, the Density Frame remained well-behaved even outside the formal regime of validity (large shear viscosity), offering a level of stability that makes it particularly attractive for real-world simulations. 
The numerical results from QCD kinetics for one-dimensional fluid flows are novel in their own right and  will be presented in greater detail elsewhere~\cite{MazeliauskasZhou}. 

We also compared the Density Frame against two other formulations of viscous fluid dynamics: BDNK and DNMR.
Within the regime of applicability of hydrodynamics, Density Frame and BDNK showed an appropriate level of agreement. However, at least with the numerical scheme of \cite{Pandya:2021ief}, the
BDNK approach was markedly less robust at the boundary of applicability, especially in scenarios involving strongly relativistic flows. 
Comparisons were also made with the DNMR second-order hydrodynamics as implemented in the computer code MUSIC~\cite{Schenke:2010nt, Schenke:2010rr, Paquet:2015lta}. Using QCD kinetics as an absolute standard, the Density Frame produced results that were competitive with MUSIC, demonstrating its potential as an alternative approach. 

From a theoretical perspective the Density Frame is also appealing, representing relativistic viscous hydrodynamics in its purest form. Indeed, the Density Frame is the only numerically stable formulation of viscous hydrodynamics that has no additional parameters beyond the shear and bulk viscosities and the equation of state. Moreover, the Density Frame has no ``non-hydrodynamic" modes,   and thus will serve as an attractive foil to previous simulations and corresponding fits to heavy-ion data.  If the extracted shear and bulk viscosities of the QGP are insensitive to non-hydrodynamic modes, the Density Frame is expected to yield similar results  with fewer parameters. The differences between the fits will bound the theoretical uncertainty in the extracted QGP parameters. 

Another compelling feature of the Density Frame is that it fits nicely into the framework of dissipative stochastic processes~\cite{FoxUhlenbeck,Generic}. Indeed, the numerical updates in the Density Frame are simple -- combining a symplectic step of ideal hydrodynamics with a viscous step that is similar to a diffusion equation.
%
In a companion paper~\cite{TheoryPaper}, we proposed a formulation of stochastic hydrodynamics with the Density Frame which replaces the diffusive step with Metropolis updates. These stochastic updates consistently incorporate thermal noise. Implementing  the proposed Metropolis algorithm is a clear next step,  and it is a step which is well motivated both by the results presented here and by our earlier work~\cite{Basar:2024qxd}. 

In summary, the Density Frame represents a promising and theoretically elegant approach to relativistic viscous hydrodynamics. Its stability, minimal number of parameters, and compatibility with stochastic processes make it a valuable tool for both theoretical exploration and practical simulations.

\begin{acknowledgments}
  We gratefully acknowledge  A.~Pandya and F.~Pretorius who provided the computer code for the finite difference BDNK simulations used in this work. 
J.B. and D.T. are supported by the U.S. Department of Energy, Office of Science, Office of Nuclear Physics, grant No. DE-FG-02-88ER40388.      R.S. is supported partly by a postdoctoral fellowship of West University of Timișoara, Romania.
M.~S.~and J.-F.~P.~are supported by Vanderbilt University and by the U.S. Department of Energy, Office of Science under Award Number DE-SC-0024347.
A.M. and F.Z. acknowledge support by the DFG through Emmy Noether Programme (project number 496831614) and CRC 1225 ISOQUANT (project number 27381115) as well as grant no INST 39/963-1 FUGG (bwForCluster NEMO).
\end{acknowledgments}

\appendix

\section{The Density Frame from the Landau Frame}
\label{app:dfdetails}

In this section we detail the steps leading to \Eq{eq:txx1d} and \eqref{eq:noisekernelintro}.
This is an elaboration of the paragraph surrounding \Eq{eq:Landauedense}.  The presentation is self-contained, but limited to $1+1$ dimensional motion and fluids with vanishing bulk viscosity.  The general case is treated in our companion paper~\cite{TheoryPaper}.

Below we will need the derivatives of the ideal stress tensor
 \st
  \label{eq:framechange2}
  {\mathcal X}^{\mu\nu\rho} \equiv\frac{\partial \tideal^{\mu\nu}}{\partial \beta_{\rho} }
  =\frac{e+p}{\beta} \left[ \frac{1}{c_s^2} u^{\mu} u^{\nu} u^{\rho} + (u^{\mu} \Delta^{\nu\rho}  + u^{\nu} \Delta^{\mu\rho} + u^{\rho} \Delta^{\mu\nu})  \right]  \, .
 \stp
 The thermodynamic susceptibility is  $\chi^{\mu\nu} \equiv {\mathcal X}^{0\mu\nu}$ and describes the equilibrium fluctuations of the densities $\tideal^{0\mu}$. The 
inverse susceptibility is
 \begin{align}
   \chi^{-1}_{\rho\sigma}  =& \frac{\beta}{(e  + p)\gamma }  \left[  \frac{c_s^2}{1 - c_s^2 v^2} \,   \left(u_\rho + \frac{1}{\gamma} \Delta^{0}_\rho \right)  \left(u_\sigma + \frac{1}{\gamma} \Delta^{0}_\sigma \right)  + \Delta_{\rho \sigma} \right ]   \, ,
\end{align} 
and describes the equilibrium fluctuations of the corresponding Lagrange multipliers $\beta_{\mu}$. 

Now to change frames from the Landau Frame to the 
Density Frame,  we shift the four velocity, $\ubeta_{\mu}=\beta_{\mu} + \delta\beta_{\mu}$,   to
cancel the temporal components of the viscous stress, $\underline{\Pi}^{0\nu}$
  \begin{align}
  T^{0\mu} =&  \tideal^{0\mu}(\beta + \delta \beta) + \underline{\Pi}^{0\mu} \equiv \tideal^{0\mu}(\beta)  \, .
\end{align}
To first order we find $\delta\beta_{\mu} =-\chi^{-1}_{\mu\nu} \Pi^{0\nu}$.   The shift propagates into the spatial stress
\begin{align}
  T^{xx} = \tideal^{xx}(\beta + \delta \beta)  + \underline{\Pi} ^{xx}
  \simeq&
  \tideal^{xx}(\beta)  +  \left(  \underline{\Pi}^{xx}- \frac{\partial \tideal^{xx}}{\partial \tideal^{0\nu}} \Pi^{0\nu} \right)  \, . 
\end{align}
In approximating the stress we expanded in $\delta\beta_{\mu}$ and used
\st
\label{eq:derivs}
\frac{\partial \tideal^{xx}  }{\partial \tideal^{0\nu}} \equiv \frac{\partial \tideal^{xx}(\beta) }{\partial \beta_\mu} \, \chi^{-1}_{\mu\nu} =  
\left(
 \frac{c_s^2 - v^2}{1 - c^2_s v^2} \, , \,  \frac{ 2\left(1- c^2_s\right) v}{1-c^2_s v^2} \right) \, .
\stp
The frame transformation can be written in a matrix notation
\st
\label{eq:frametrans}
T^{xx}  = \tideal^{xx}(\beta)  +  \kappa^{xx}_{\mu\nu}\,  \underline{\Pi}^{\mu\nu}\, ,  \qquad \mbox{with} \qquad \kappa^{xx}_{\mu\nu}\,    \equiv  \delta^{x}_{(\mu} \delta^{x}_{\nu)}  - \frac{\partial \tideal^{xx}  }{\partial \tideal^{0\rho}}  \delta^{0}_{(\mu} \delta^{\rho}_{\nu)}   \, .
\stp

The viscous strain in $\underline{\Pi}^{\mu\nu}$ given  in \eqref{eq:Landaustress} has time derivatives. These are approximated with spatial derivatives using the ideal equations of motion:
\st
\partial_{\nu}\tideal^{\mu\nu} = {\mathcal X}^{\mu\nu\rho} \partial_{(\nu}\beta_{\rho)} = 0 \,. 
\stp
Extracting the time derivatives we find 
\begin{align}
\partial_{(0} \beta_{\mu)} =& -\frac{\partial \tideal^{xx} }{\partial \tideal^{0\mu } } \partial_x \beta_x   \, ,
\end{align}
and thus we have an approximation for the strains in the fluid based on spatial derivatives
\st
\label{eq:betaapprox}
\partial_{(\mu} \beta_{\nu)} \simeq \kappa_{\mu\nu}^{xx}\,  \partial_{(x}\beta_{x)} \, .
\stp

With the use of frame transformation in \eqref{eq:frametrans}, the Landau stress in \eqref{eq:Landaustress}, the approximation in \eqref{eq:betaapprox}, and the explicit derivatives in \eqref{eq:derivs},   we find 
\begin{align}
  T^{xx} \simeq& 
  \tideal^{xx}(\beta)  - 2T \eta \, \kappa_{\mu\nu}^{xx} \left(\Delta^{(\mu\rho} \Delta^{\nu) \sigma} - \frac{1}{3} \Delta^{\mu\nu} \Delta^{\rho\sigma} \right) \kappa_{\rho\sigma}^{xx} \, \partial_{(x}\beta_{x)}  \, ,  \\
  =& \tideal^{xx}(\beta)  - T \kappa^{xxxx}(v) \, \partial_{(x}\beta_{x)} \, ,
\end{align}
where $\kappa^{xxxx}(v)$ is given in \eqref{eq:noisekernelintro}.


\section{Comparison of numerical algorithms}
\label{app:numericaldetails}

In \Sect{sec:numericsoverview} we discussed several approaches to simulating viscous hydrodynamics in the density
frame before settling on the linearly implicit Extrapolated IMEX (EIMEX) scheme as our baseline method.
In this appendix  we will compare the IMEX Runge-Kutta scheme  to this method. 

First, we will describe the EIMEX scheme in greater detail. The update given in \eq{eq:FofVschematic} defines
a base step.  A base step solution over a time  $\Delta t =H$  is  denoted $T_{1, 1}$  and is accurate to 
first order in $H$. Taking two base steps of size $H/2$  yields an estimate $T_{2, 1}$ for the solution which is also
accurate to first order in $H$. (The first index is the number of steps and the second index is 
the order.)
However, combining the information from the $T_{1,1}$  and $T_{2,1}$ yields an estimate which 
is accurate to order in $H^2$ inclusive and constitutes the complete step at second order
\st
  T_{2,2} = T_{2, 1} + (T_{2, 1} - T_{1,1}) \, .
\stp
At third order in accuracy, an extrapolation scheme based on $T_{3,1}$ (which makes three steps of size $H/3$) and $T_{2,1}$ and $T_{1,1}$ 
can be readily constructed~\cite{Constantinescu}. The numerical results presented in 
\Sect{sec:numerics} are based on this third order algorithm. 

Next we will describe the  IMEX schemes.
As discussed in \Sect{sec:numericsoverview}, in IMEX Runge-Kutta schemes one
divides the full evolution into an explicit ideal step and an implicit
viscous step. The implicit viscous step involves solving the non-linear
equations, \eq{eq:FofV} and \eq{eq:FofVschematic},  given in the body of the paper.  
In a simple IMEX Euler Runge-Kutta the explicit and implicit parts are treated
sequentially. A first order two stage  Euler scheme ({\tt ARS(1,2,1) }  from 
\cite{ASCHER1997151}) involves solving the non-linear equation
\begin{subequations}
\begin{align}
  \label{eq:stages_euler}
  V^{(1)} =& V^n \,,\\
  V^{(2)} =& V^n + \Delta t \left( f(V^{(1)})  +  g(V^{(2)}) \right)\,,
\end{align}
\end{subequations}
before finally completing the step
\st
\label{eq:completion_euler}
    V^{n+1} =   V^n + \Delta t  \left( f(V^{(2)})  +  g(V^{(2)}) \right) \, .
\stp

Higher order temporal accuracy can be achieved using a variety of
methods~\cite{ASCHER1997151}. We used the 
ARK3 scheme of Kennedy and  Carpenter~\cite{KENNEDY2003139} as implemented in PETSc. The stages
take a form analogous to  \Eq{eq:stages_euler}
\st
V^{(i)} =  V^n +  \Delta t  \sum_{j=1}^{i-1} a_{ij}^E f(V^{(j)}) + \Delta t \sum_{j=1}^i a_{ij}^I g(V^{(j)})\,,
\stp
which involves solving nonlinear equations form $V^{(i)} = V_{*} + k \Delta t g(V^{(i)})$  with a Newton method
in each stage. Each stage provides approximation to the solution  at time $t^{(i)}= t_n + c_i \Delta t$. Finally the 
step is completed with an update analogous to \Eq{eq:completion_euler}
\st
V^{n+1 } = V^n + \Delta t \sum_{j=1}^s b_i \left( f(V^{(j)}) + g(V^{(j)}) \right).
\stp
A table of the required coefficients for the {\tt ARK3(2)L[2]SA} scheme  is given Appendix C of  \cite{KENNEDY2003139}.

\begin{figure}
  \centering
  \begin{minipage}[c]{0.55\textwidth}
  \includegraphics[width=1.0\textwidth]{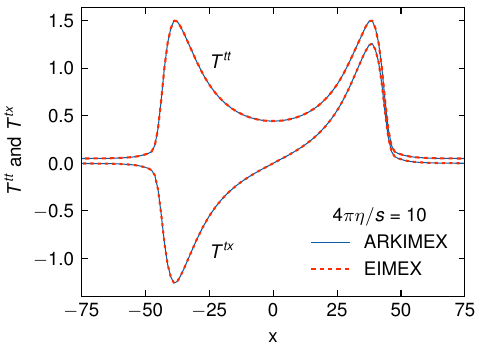}
  \end{minipage}
    \caption{A comparison between the semi-implicit EIMEX scheme used in the body of the paper
      with the implicit IMEX scheme based on the ARK3 solver at time $t=47.5\,{\rm GeV}^{-1}$. 
      The  test is {\tt test2} described in \Fig{fig:smoothtest2} and features moderately
      relativistic flow and a developing shock. The shear viscosity is relatively large -- see \Eq{eq:Knudsen}
      and surrounding text for context.
  }
  \label{fig:test2comparison}
\end{figure}

In \Fig{fig:test2comparison} we compare the solutions using the semi-implicit (EIMEX) approach 
used in the body of the text and the implicit IMEX approach based on the ARK3 stepper.  
We generally found that the solutions are quite close. However, the semi-implicit scheme is considerably faster and equally robust for the types of applications  we envision.

\section{Comparison of M\"uller-Israel-Stewart type theories}
\label{app:MUSICcomparisons}

In M\"uller-Israel-Stewart (MIS) type theories, the shear tensor $\pi^{\mu\nu}$ is elevated to be an independent variable with  its own relaxation-type evolution equation. MUSIC~\cite{Schenke:2010nt, Schenke:2010rr, Paquet:2015lta} can solve either the original M\"uller-Israel-Stewart hydrodynamics, or a version of the DNMR equations~\cite{Denicol:2012cn}.
Specifically, in absence of bulk viscosity,  MUSIC generally solves the following DNMR equation for the shear tensor $\pi^{\mu\nu}$:
\begin{equation}
    \tau_\pi\Dot{\pi}^{\langle\mu\nu\rangle} + \pi^{\mu\nu} = 2\eta\sigma^{\mu\nu} - \delta_{\pi\pi}\pi^{\mu\nu}\theta + \phi_7 \pi_\alpha^{\langle \mu}\pi^{\nu\rangle\alpha} - \tau_{\pi\pi}\pi_{\alpha}^{\langle\mu}\sigma^{\nu\rangle\alpha}.
\end{equation}
The angular brackets in the indices indicate the symmetrized and 
traceless part of a tensor. The expansion rate is $\theta = \nabla_\mu u^\mu$  where $\nabla^\mu = \Delta^{\mu\nu}\partial_\nu$ and the Navier-Stokes shear stress tensor is given by $\sigma^{\mu\nu}$,
\begin{equation}
    \sigma^{\mu\nu} = \frac{1}{2}\left[\nabla^\mu u^\nu + \nabla^\nu u^\mu - \frac{2}{3}\Delta^{\mu\nu}\theta\right].
\end{equation}

The transport coefficients $\delta_{\pi\pi}, \phi_7$ and $\tau_{\pi\pi}$ are fixed by kinetic theory \cite{Molnar:2013lta,Denicol:2014vaa}:
\begin{align*}
\delta_{\pi\pi} & =  \frac{4}{3} \tau_\pi, \\
\phi_7 & =  
\frac{9}{70} \left[ \frac{4}{5} \frac{\tau_\pi}{\eta} \right], \\
\tau_{\pi\pi} & = \frac{10}{7} \tau_\pi.
\end{align*}
The original MIS theory is recovered by setting $\phi_7 = \tau_{\pi\pi} = 0$. 

\begin{figure}
  \centering
  \begin{minipage}[c]{0.49\textwidth}
    \includegraphics[width=1.0\textwidth]{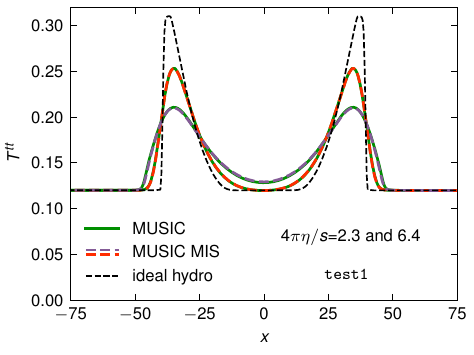}
  \end{minipage}
  \begin{minipage}[c]{0.49\textwidth}
    \includegraphics[width=1.0\textwidth]{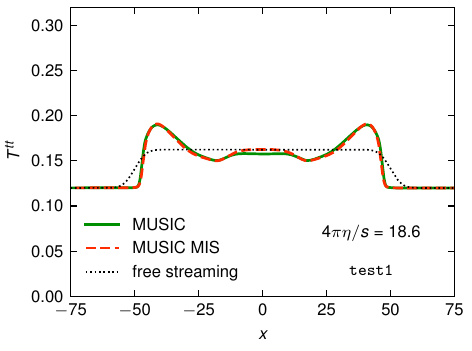}
  \end{minipage}
  \begin{minipage}[c]{0.49\textwidth}
    \includegraphics[width=1.0\textwidth]{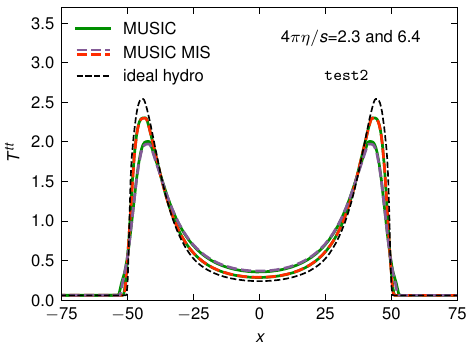}
  \end{minipage}
  \begin{minipage}[c]{0.49\textwidth}
    \includegraphics[width=1.0\textwidth]{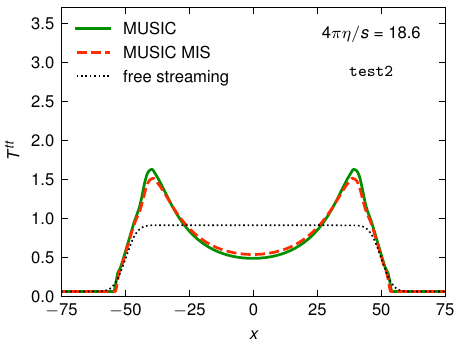}
  \end{minipage}
  \caption{The upper panels show the evolution of energy density $T^{tt}$ of {\tt test1} at $t=50\, {\rm GeV}^{-1}$ in second order hydrodynamics (MUSIC) compared with MUSIC in MIS mode for $4\pi \eta/s=2.3,6.4$ (left) and $4\pi \eta/s=18.6$ (right). The lower panels show the same for {\tt test2}. The dashed black lines on the left panel correspond to ideal hydro while the dotted black lines on the right panels correspond to free streaming ($\eta/s = \infty$).
  }
  \label{fig:MUSIC_MIS}
\end{figure}

We compare the energy density evolution for {\tt test1} and {\tt test2} for DNMR MUSIC and MIS MUSIC evolution in Fig. \ref{fig:MUSIC_MIS}. There is a fairly good agreement between the outcomes of two different second-order hydrodynamics theories. The differences arise only at large values of shear viscosities. Consequently, it is reasonable to compare Density Frame hydrodynamics with either second-order theory as a representative of the whole group.

\begin{figure}
  \centering
  \begin{minipage}[c]{0.49\textwidth}
    \includegraphics[width=1.0\textwidth]{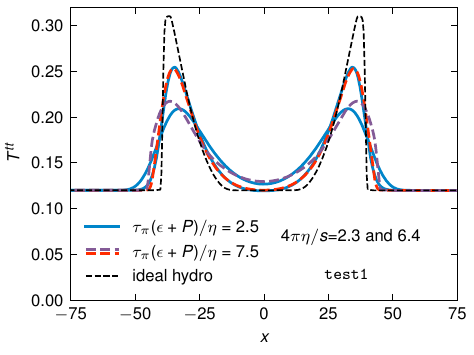}
  \end{minipage}
  \begin{minipage}[c]{0.49\textwidth}
    \includegraphics[width=1.0\textwidth]{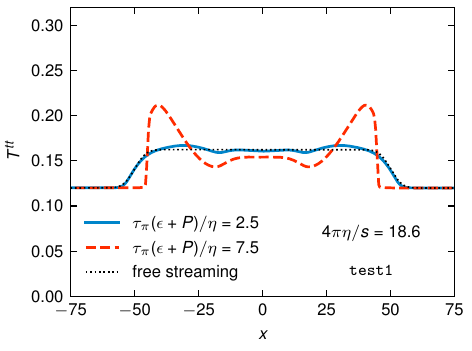}
  \end{minipage}
  \begin{minipage}[c]{0.49\textwidth}
    \includegraphics[width=1.0\textwidth]{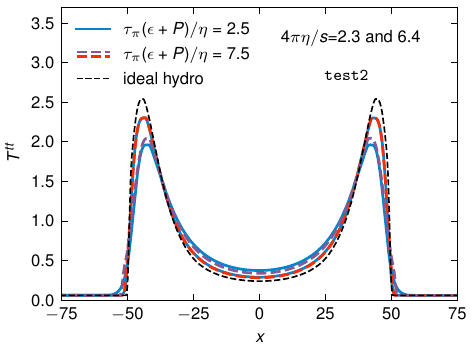}
  \end{minipage}
  \begin{minipage}[c]{0.49\textwidth}
    \includegraphics[width=1.0\textwidth]{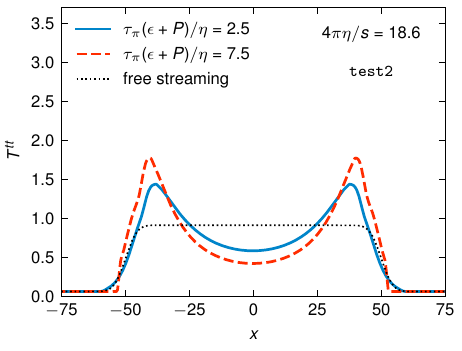}
  \end{minipage}
  \caption{The upper panels show the evolution of energy density $T^{tt}$ of {\tt test1} at $t=50\, {\rm GeV}^{-1}$ in second order hydrodynamics (MUSIC) for $\tau_\pi(\epsilon + P)/\eta = 2.5$ and $7.5$ for $4\pi \eta/s=2.3,6.4$ (left) and $4\pi \eta/s=18.6$ (right). The lower panels show the same for {\tt test2}. The dashed black lines on the left panel correspond to ideal hydro while the dotted black lines on the right panels correspond to free streaming ($\eta/s = \infty$).}
  \label{fig:MUSIC_taupi}
\end{figure}

The shear relaxation time $\tau_\pi$ is often parametrized as $\tau_\pi = b_\pi \eta/(\epsilon + P)$ where $b_\pi$ is a positive real number. For weakly-coupled QCD kinetic theory, $b_\pi$ is expected to range from 5 to 7.5 \cite{York:2008rr,Ghiglieri:2018dgf}. In MUSIC, the default value of $b_\pi$ is 5. In Fig. \ref{fig:MUSIC_taupi} we compare the energy-density evolution for $b_\pi = 2.5$ and $7.5$. For small viscosities, which is the region of applicability of these theories, the difference in energy density for different choices of $b_\pi$ is comparable to the difference between Density Frame hydrodynamics and default MUSIC. This makes Density Frame hydrodynamics a reliable alternative in this regime.

It is also interesting to note that energy density spills beyond the causality edge in second-order hydrodynamics for a small enough value of $\tau_\pi$.

%

\end{document}